\documentclass{article}
\usepackage{spconf,amsmath,graphicx}
\usepackage{color}
\usepackage{url}
\usepackage{booktabs}
\usepackage[fontsize=9.15pt]{fontsize}
\usepackage{kotex}
\usepackage{comment}
\usepackage{multirow}
\usepackage[ruled,vlined]{algorithm2e}

\title{PSEUDO-LABEL TRANSFER FROM FRAME-LEVEL TO NOTE-LEVEL IN A TEACHER-STUDENT FRAMEWORK FOR SINGING TRANSCRIPTION FROM POLYPHONIC MUSIC}
\name{Sangeun Kum$^{1}$ \hspace{.65cm} Jongpil Lee$^{1}$ \hspace{.65cm} Keunhyoung Luke Kim$^{1}$ \hspace{.65cm} Taehyoung Kim$^{1}$ \hspace{.65cm} Juhan Nam$^{1,2}$}

\address{$^1$Neutune Research, Seoul, South Korea\\
$^2$Graduate School of Culture Technology, KAIST, Daejeon, South Korea}

\begin{document}
%
\maketitle
\begin{abstract}
Lack of large-scale note-level labeled data is the major obstacle to singing transcription from polyphonic music. We address the issue by using pseudo labels from vocal pitch estimation models given unlabeled data. The proposed method first converts the frame-level pseudo labels to note-level through pitch and rhythm quantization steps. Then, it further improves the label quality through self-training in a teacher-student framework. To validate the method, we conduct various experiment settings by investigating two vocal pitch estimation models as pseudo-label generators, two setups of teacher-student frameworks, and the number of iterations in self-training. The results show that the proposed method can effectively leverage large-scale unlabeled audio data and self-training with the noisy student model helps to improve performance. Finally, we show that the model trained with only unlabeled data has comparable performance to previous works and the model trained with additional labeled data achieves higher accuracy than the model trained with only labeled data.

\end{abstract}
\section{Introduction}
\label{sec:intro}


The goal of singing transcription from polyphonic music (STP) is to transcribe the monophonic vocal in polyphonic music into musical notes denoted by onset, offset, and the score pitch. STP includes several sub-tasks such as singing voice detection, pitch estimation, note-level segmentation, and onset/offset detection \cite{nishikimi2019automatic}. It is a challenging task due to the high variability of singing voice in terms of timbre, expressions or formant modulation. Furthermore, multiple instrument sources in polyphonic music make the task even harder. Therefore, many of the singing transcription methods have focused on monophonic vocal sources \cite{yang2017probabilistic,nishikimi2019automatic} or have employed pre-trained source separation models as pre-processing to extract vocal sources from polyphonic music \cite{hsu2021vocano, wang2021preparation}.

Although we can leverage the power of a deep neural network in supervised settings for STP, the major obstacle is the lack of large-scale note-level label data. While there are several datasets released for singing transcription, they cover only monophonic singing voice \cite{gomez2013towards,molina2014evaluation} or have low quality as the labeling was automated without manual refinement \cite{wang2021preparation,Meseguer2018dali}. The lack of note-level label data can be addressed by additional singing voices that sung the song \cite{zhu2017fusing} but it requires more resources and has an issue of data distribution. A recent work released a dataset of human-annotated note-level labels for 500 Chinese songs \cite{wang2021preparation}. While this is highly beneficial for STP research, such manual note-level labeling is time-exhausting and expensive. 

We address the lack of labeled data by using pseudo labels from vocal pitch estimation models.
Vocal pitch estimation or melody extraction is a sub-task of STP that estimates the frame-level instantaneous pitch of vocal source music \cite{salamon2014melody}. Recently, deep neural network models for the task have been extensively studied, showing superior performances \cite{bittner2017deep,lu2018vocal,kim2018crepe, hsieh2019streamlined,kum2019joint}. Using the vocal pitch estimation models, we first convert the frame-level pitch contours to note-level piano roll as pseudo labels. We then use the the note-level pseudo labels for self-training of a STP model in a teacher-student framework. Along with the noisy student model that employs various data augmentation techniques in a teacher-student framework \cite{kum2020semi}, we show that the newly trained neural network model consistently improves the performance for STP. We validate the proposed method in various experiment setups using the Cmedia and MIR-ST500 datasets. Finally, we show that the model trained with only unlabeled data has reasonable performances compared to previous works and, the model trained with additional labeled data, achieves higher accuracy than the model trained with only labeled data. 



\section{Method}
 
The proposed method is composed of two stages. The first stage extracts F0 pitch contours from vocal pitch estimation models and converts them to note-level pseudo labels. The second stage uses the note-level pseudo labels to train a new neural network for STP in a teacher-student framework. The details of each stage are explained below.  



\subsection{Pseudo Labels: Frame-level to Note-level}
\label{ssec:note_segmenation}

\begin{figure}[t]
\begin{minipage}[b]{\columnwidth}
  \centering
  \centerline{\includegraphics[scale=0.7]{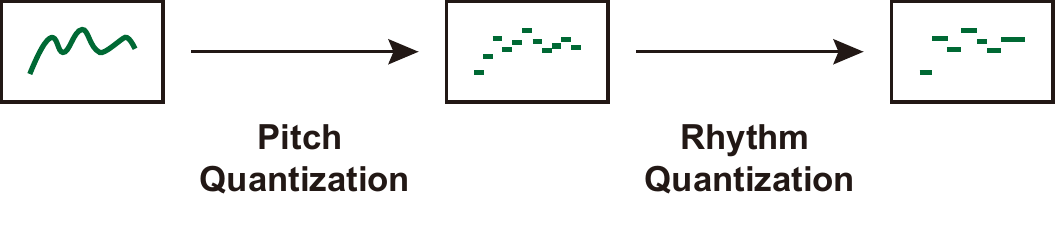}}
  \centerline{(a) Conversion from pitch contours to piano rolls }\medskip\medskip\medskip
\end{minipage}
\begin{minipage}[b]{\columnwidth}
  \centering
  \centerline{\includegraphics[scale=0.9]{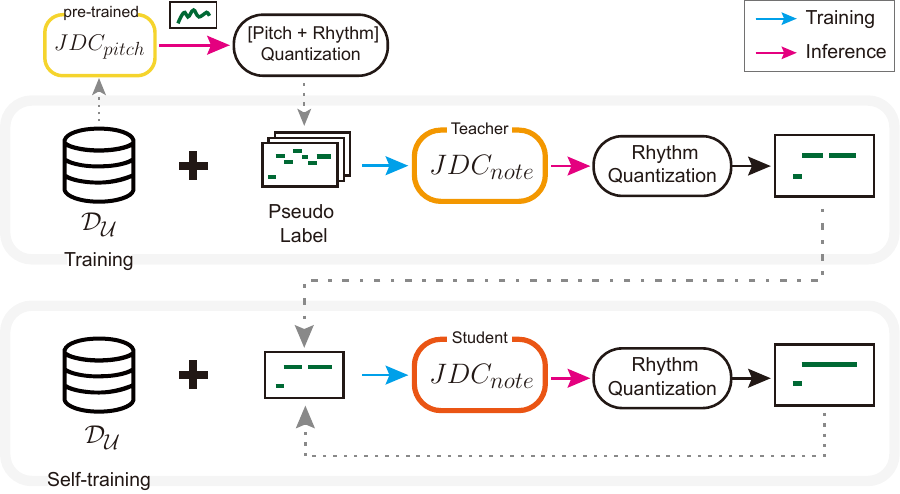}}
  \centerline{(b) The proposed teacher-student framework. }\medskip\medskip
\end{minipage}
\caption{Illustrations of pseudo label transfer from frame-level to note-level in the teacher-student framework. $\mathcal{D_U}$ is a large-scale unlabeled audio data, $JDC_{pitch}$ is a pretrained vocal pitch estimation model, and $JDC_{note}$ is a STP model trained with note-level pseudo labels.}
\label{fig:fig_1}
\end{figure}

Figure \ref{fig:fig_1} (a) illustrates the conversion steps from frame-level pitch contours to note-level piano rolls. The pitch contours are obtained from a vocal pitch estimation model given unlabeled audio data (polyphonic music with vocals). 
The first step is pitch quantization which rounds the continuous pitch to semi-tone steps. In many pitch estimation models based on neural networks, the output is represented as the softmax function over quantized pitch values where the neighboring pitches are much smaller than one semi-tone \cite{kim2018crepe,kum2019joint}. We take the pitch with the highest confidence and quantize it to the nearest MIDI note number.
The second step is rhythm quantization. This processing ``snaps'' the fragments of the quantized pitch lines to beat-based units. It is carried out by smoothing the quantized pitch with a series of three median filters. In order to make the filtered outputs beat-based units, we set the sizes of the median filters to 1/32, 1/16, and 1/12 beat, respectively, given a tempo. We estimated the tempo for each audio track using the function in librosa \cite{mcfee2015librosa}. We observed that the three cascaded filters progressively improved the label quality. In addition, we removed small fragments that were too short to be considered as singing notes. In the experiment, we set the threshold to 1/16 beat. Finally, we added a simple rule to minimize octave errors. 

\subsection{Self-Training in the Teacher-Student Framework}
\label{ssec:TS_framework}

Figure \ref{fig:fig_1} illustrates the proposed teacher-student framework. We first build a new neural network model for STP and train it using the note-level pseudo labels obtained from the first stage. We then iterate the training using the same configuration of neural network in the teacher-student framework. This self-training in a teacher-student framework has been studied in diverse domains such as image classification~\cite{berthelot2019mixmatch,xie2020self}, speech recognition~\cite{movsner2019improving}, and audio classification~\cite{lu2019semi}. In particular, random data augmentation is a key element to improve the model performance. In our method, we primarily used the noisy student model where the student network,  takes unlabeled data that is perturbed by the random audio augmentation while the teacher network takes the original input to generate the pseudo-labels \cite{kum2020semi}. For random audio augmentation, multiple audio effects are randomly applied such as overdrive, reverb, audio equalizer, and audio filter to manipulate the audio. In addition, between the teacher network and the student network, we add the rhythm quantization as a post-processing to enhance the transcription performance. For applying the median filtering and the following rules in the rhythm quantization, we use one-hot vectors (i.e., hard label) as pseudo labels. In the mean time, we compare the noisy student model to the basic teacher-student model which does not use the random audio augmentation in the experiment.

\subsection{Model Architecture}
\label{ssec:model}

\begin{figure}[t]
\centering
\includegraphics[scale=0.8]{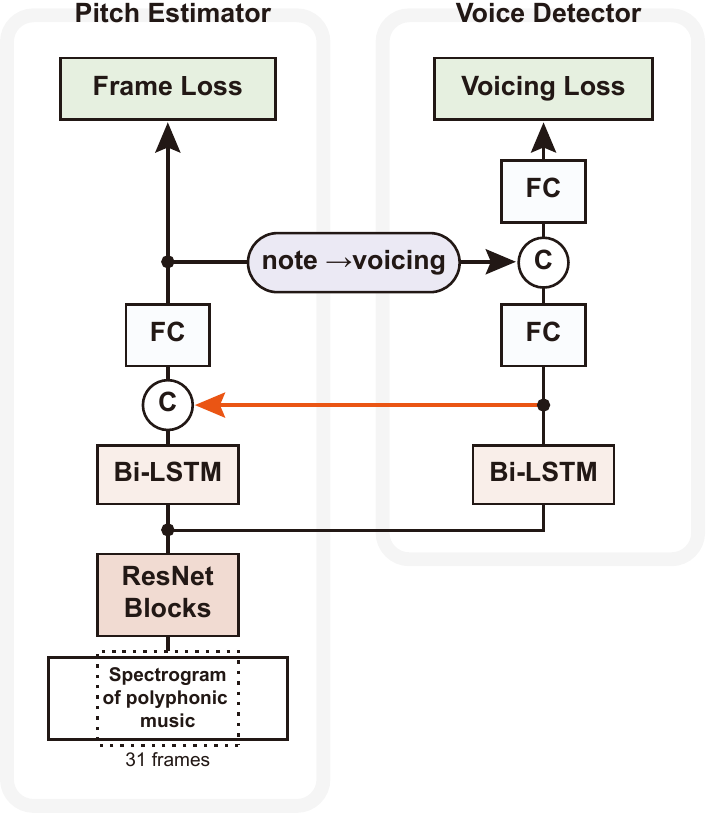}
\caption{The model architecture for $JDC_{note}$. ``C'' indicates feature concatenation.}
\label{fig:model_architecture}
\end{figure}
\vspace{-3mm}

The model architecture for STP is based on the joint detection and classification (JDC) model, which which is originally proposed for vocal melody extraction \cite{kum2019joint}. The JDC model is a convolutional recurrent neural network (CRNN) with two outputs where one detects the presence of a singing voice and the other estimates the pitch. We modified the JDC model in two folds. First, we set the softmax output of the pitch estimator to have the semi-tone resolution. Second, we added a connection from the output of Bi-LSTM in the voice detector to the pitch estimator as seen in Figure \ref{fig:model_architecture}. The STP model is denoted as JDC$_{note}$. In our preliminary study, we observed that the modified JDC model performs better than the original one in STP (the average COnPOff is 2.25\% higher on the MIR-ST500 and Cmedia test sets). In the experiments, we also use the original JDC model as a pretrained vocal pitch estimator. The model is denoted as JDC$_{pitch}$ to distinguish it from JDC$_{note}$.

\section{Experiments}

\subsection{Datasets}
We used various labeled or unlabeled datasets that contain singing voice for training or testing the STP model. 

\subsubsection{Training data}
We used a large-scale unlabeled datasets ($\mathcal{D_U}$) to generate pseudo labels in the proposed method. They include two public datasets (DSD100 \cite{SiSEC16} and Free Music Archive (FMA) \cite{fma_dataset}) and an in-house dataset. As to FMA, we used fma\_large, a subset of FMAs with up to 106,574 tracks. Since it contains numerous non-vocal tracks covering a variety of genres and sounds, we mitigated the data imbalance between vocal and non-vocal examples by selecting only vocal tracks as in~\cite{kum2020semi}. The in-house dataset contains 2000 K-pop songs crawled from YouTube. We also used a labeled dataset for STP for the ablation study. MIR-ST500 \cite{wang2021preparation} consists of 500 Chinese pop songs and human-labeled notes that correspond to the vocal melody. Among them, we used the official split of 400 songs as a training set. 



\subsubsection{Test data}
We evaluated the models mainly with two public test sets, Cmedia and MIR-ST500 (the test split). The Cmedia dataset provides note-level labels and YouTube links to download audio files. It has been used in the Music Information Retrieval Evaluation eXchange (MIREX 2020) \footnote{\url{https://www.music-ir.org/mirex/wiki/2020:Singing_Transcription_from_Polyphonic_Music}}. It contains 100 Chinese pop songs. We used it only for testing in this experiment to compare performance with other models. For MIR-ST500, we used the official split of 100 songs as a test set. 


\subsection{Training Details}

The model configuration of the CRNN architecture in JDC$_{note}$ is almost the same as  JDC$_{pitch}$ in \cite{kum2019joint}. We trained the JDC$_{note}$ using the Adam optimizer for 100 epochs with a learning rate of 0.003, batch size of 64 on 1 GPU. We used a learning rate schedule that reduces the learning rate by 0.7 times if validation accuracy did not increase within 9 epochs. The model and the training procedures were implemented using Keras with TensorFlow 2.3. The source code of the proposed method and the models are available at \url{https://github.com/keums/icassp2022-vocal-transcription}.



\begin{table}[!t]
\centering
\resizebox{0.9\columnwidth}{!}{%
\begin{tabular}{@{}ccccc@{}}
\toprule
 &
  \multicolumn{2}{c}{\textbf{Initial Pseudo Labels}} &
  \multicolumn{2}{c}{\textbf{JDC$_{note}$}} \\ \midrule
\textbf{\begin{tabular}[c]{@{}c@{}}Repurposed\\  Models\end{tabular}} &
  \begin{tabular}[c]{@{}c@{}}Demucs\\ + CREPE\end{tabular} &
  JDC$_{pitch}$ &
  \begin{tabular}[c]{@{}c@{}}Demucs\\ + CREPE\end{tabular} &
  JDC$_{pitch}$ \\ \midrule
\textbf{COnPOff} & 22.43      & 25.44      & 24.71        & 28.97      \\
\textbf{COnP}    & 45.01      & 48.48      & 48.64        & 53.32      \\
\textbf{COn}     & 57.65      & 61.94      & 62.32        & 64.74      \\ \bottomrule
\end{tabular}%
}
\caption{Comparison of pitch estimation models for their initial note-level pseudo labels (Initial Pseudo Labels) and the predicted results from JDC$_{note}$ trained with the initial pseudo labels. They were all evaluated on Cmedia.}
\label{table:exp_1}
\end{table}



\begin{table}[!t]
\centering
\resizebox{0.7\columnwidth}{!}{%
\begin{tabular}{ccccc}
\hline
                 & \multicolumn{2}{c}{\textbf{Cmedia}} & \multicolumn{2}{c}{\textbf{MIR-ST500}} \\ \hline
\textbf{Models}  & TS             & NS                 & TS              & NS                   \\ \hline
\textbf{COnPOff} & 28.97	      & 29.62	           & 22.12	         & 22.62                \\
\textbf{COnP}    & 53.32	      & 54.55	           & 40.01	         & 40.70                \\
\textbf{COn}     & 64.74	      & 65.61	           & 56.90	         & 57.87                \\ \hline
\end{tabular}
}%
\caption{Comparison of the basic teacher-student model (TS) and noisy student (NS) model.}
\label{table:exp_2}
\end{table}

\section{Ablation Study}

We conducted an ablation study to verify the proposed method for STP. In the following experiments, we evaluated STP performance using the metrics proposed in~\cite{molina2014evaluation}: COn (Correct Onset), COnP (Correct Onset and Pitch), and COnPOff (Correct Onset, Pitch and Offset). We set the onset threshold to 50 msec, and the offset threshold to 50 msec or (0.2$\times$note duration), whichever is higher, following the previous works. We computed the average F1-score for the metrics using \textit{mir\_eval}~\cite{raffel2014mir_eval}. 



\subsection{Comparison of Pitch Estimation Models}
\label{sec:Comparison_of_Pitch_Estimation_Models}

The goal of this experiment is to evaluate the efficacy of the repurposed neural network models that predict vocal pitch contours as a note-level pseudo-label generator in our method. While we primarily used the pretrained JDC$_{pitch}$ \cite{kum2019joint}\footnote{The JDC$_{pitch}$ model was trained with several datasets with pitch labels, which were not used in this paper.} for vocal pitch estimation in the experiments, we also compared it to a combination of Demucs \cite{defossez2019demucs} and CREPE \cite{kim2018crepe} because they are widely used pretrained models for sound source separation and monophonic pitch estimation, respectively. Specifically, we separated out vocal audio from the Demucs outputs and fed them to CREPE to estimate frame-level pitch. We converted the frame-level pitch contours to note-level pseudo labels based on the method in Section \ref{ssec:note_segmenation}. We first evaluated the accuracy of the note-level pseudo labels and show the result on the left side of Table \ref{table:exp_1}. While Demucs and CREPE are models with high performance, JDC$_{pitch}$ achieves 3 to 4\% higher accuracy than the combination in the three metrics. This is presumably because the separated vocal stem from Demucs includes multiple vocal sources (e.g., chorus ensembles) and CREPE predicts discontinuous pitch contours switching between different voices. On the other hand, JDC$_{pitch}$ is more robust to multiple vocal sources because it was trained to extract the main vocal melody from polyphonic music. As a next step, we trained a new neural model, JDC$_{note}$, using the note-level pseudo labels and evaluated the performance to confirm the efficacy of the repurposed neural network models. The right side of Table \ref{table:exp_1} shows that JDC$_{pitch}$ still performs better, having more accuracy gaps. Therefore, we used the note-level pseudo labels from JDC$_{pitch}$ for the remaining experiments.

\subsection{Basic Teacher-Student VS. Noisy Student}
\label{sec:TS_VS_NS}
In the experience above, we used the note-level pseudo labels from pitch estimation models to train JDC$_{note}$. The initial pseudo labels can be regarded as ``the initial teacher'' of JDC$_{note}$ in the view of teacher-student framework. In particular, we used the basic teacher-student model which does not use the random audio augmentation \cite{kum2020semi}. In this experiment, we compare the basic teacher-student (TS) model to the noisy student (NS) model where we randomly augment the input audio as described in Section \ref{ssec:TS_framework}. Table \ref{table:exp_2} shows the performances of two teacher-student models. The result shows that The NS model achieves consistently higher accuracy in all metrics than the TS model, although the gap is somewhat small in MIR-ST500. This validates that the NS model is an effective approach in the teacher-student framework as concluded in \cite{kum2020semi}. 




\subsection{Iteration of Self-training}
\label{sec:Iteration_of_Self-training}

\begin{figure}[!t]
\begin{minipage}[b]{1.0\linewidth}
  \centering
  \centerline{\includegraphics[width=8.5cm]{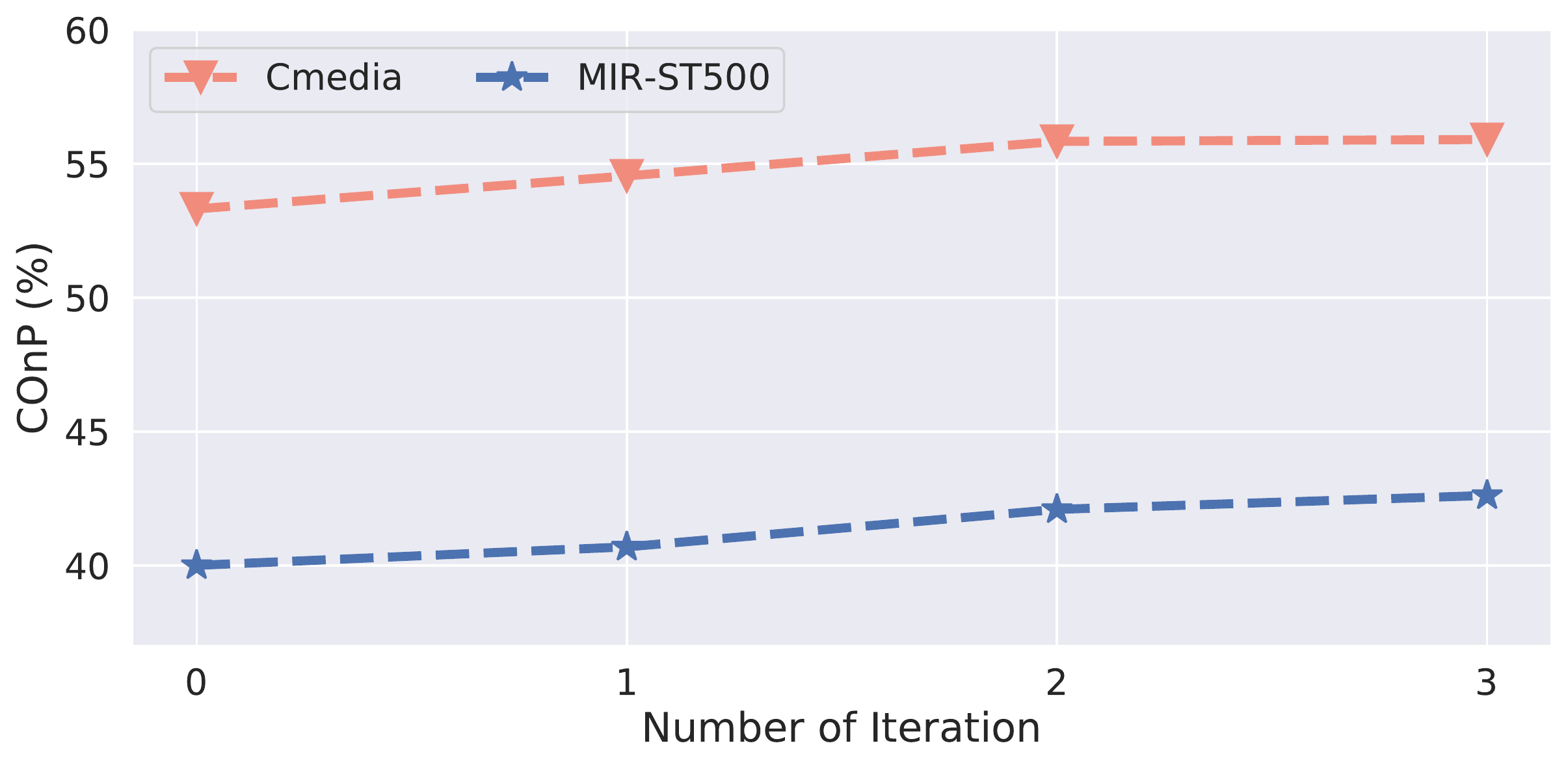}}
\end{minipage}
\caption{Effect of iterative training in the noisy student model on Cmedia and MIR-ST500.}
\label{fig:iter}
\end{figure}

We also investigated the number of self-training iterations in the NS model. We conducted up to 3 iterations of self-training for JDC$_{note}$. Figure \ref{fig:iter} shows the results for the COnP accuracy on MIR-ST500 and Cmedia. The iteration number 0 indicates when the note-level pseudo labels converted from estimated pitch contours are evaluated in Section \ref{sec:Comparison_of_Pitch_Estimation_Models}. The iteration number 1 indicates when JDC$_{note}$ is trained for the first time with the NS model in Section \ref{sec:TS_VS_NS}. In the next two iterations, the performance of JDC$_{note}$ consistently increases although the slopes become saturated. We observed similar trends in the COn and COnPOff metrics. This result indicates that the iterative training using the NS model can reliably improve the model performance. 

\subsection{Comparison with Supervised and Semi-Supervised Models}
While the proposed method in this paper aims to use only unlabeled audio data and achieve meaningful STP performance using pseudo labels from pretrained neural networks, we can also use labeled data together to further improve the performance. Table \ref{table:model_description} describes three different versions of JDC$_{note}$ trained with different sets of data. JDC$_{note}$(U) is the unsupervised model with 3 iterations of self-training from Section \ref{sec:Iteration_of_Self-training}. JDC$_{note}$(L) is a solely supervised model trained with labeled audio data. JDC$_{note}$(L+U) is a semi-supervised model trained with both the unlabeled and labeled data. The right side of Table \ref{table:sota} shows the results. The supervised model achieves significantly higher accuracy than the unsupervised model on both test sets. However, the semi-supervised model outperforms the solely supervised model. This indicates that the proposed method has synergy with supervised learning.    

\section{Comparison with Previous Works}





\begin{table}[!t]
\centering
\resizebox{0.9\columnwidth}{!}{%
\begin{tabular}{@{}ll@{}}
\toprule
 \multicolumn{2}{c}{\textbf{Description}}    \\ \midrule
\textbf{JDC$_{note}$(U)}   & Unsupervised model with unlabeled data $\mathcal{D_U}$ \\
\textbf{JDC$_{note}$(L)}   & Supervised model with labeled data $\mathcal{D_L}$   \\
\textbf{JDC$_{note}$(L+U)} & Semi-supervised model with $\mathcal{D_L}$ and $\mathcal{D_U}$   \\ \bottomrule
\end{tabular}%
}
\caption{Description of training setting}
\label{table:model_description}
\end{table}

\begin{table}[t]
\centering
\resizebox{0.9\columnwidth}{!}{%
\begin{tabular}{@{}ccccccc@{}}
\toprule
 \multicolumn{7}{c}{\textbf{Cmedia}}       \\ \midrule
\multirow{2}{*}{\textbf{Model}} & \multirow{2}{*}{HZ} & \multirow{2}{*}{VOCANO} & \multicolumn{1}{c}{\multirow{2}{*}{EFN}} & \multicolumn{3}{c}{JDC$_{note}$} \\ \cmidrule(l){5-7} 
              &  &  & \multicolumn{1}{c}{} & (U) & (L) & (L+U) \\ \midrule
\textbf{COnPOff}    & 17.18 & 28.28    & \multicolumn{1}{c}{35.13} &  30.13& 35.95 & \textbf{40.20}    \\
\textbf{COnP}       & 41.43 & 48.33    & \multicolumn{1}{c}{60.77}      &  55.84& 62.50 & \textbf{66.11}     \\
\textbf{COn}        & 63.63 & 64.56    & \multicolumn{1}{c}{\textbf{76.40}}      &  65.72& 73.88 & 75.97     \\ \midrule  \midrule

 \multicolumn{7}{c}{\textbf{MIR-ST500}}       \\ \midrule
\multirow{2}{*}{\textbf{Model}} & \multirow{2}{*}{HZ} & \multirow{2}{*}{VOCANO} & \multicolumn{1}{c}{\multirow{2}{*}{EFN}} & \multicolumn{3}{c}{JDC$_{note}$} \\ \cmidrule(l){5-7} 
              &  &  & \multicolumn{1}{c}{} & (U) & (L) & (L+U) \\ \midrule
\textbf{COnPOff}    & -     & -       & \multicolumn{1}{c}{\textbf{45.78}} &23.48  & 40.57 & 42.23           \\
\textbf{COnP}       & -     & -       & \multicolumn{1}{c}{66.63 }         &42.10  & 67.55 & \textbf{69.74}    \\
\textbf{COn}        & -     & -       & \multicolumn{1}{c}{75.44}          &58.61  & 74.94 & \textbf{76.18}   \\ \bottomrule
\end{tabular}%
}
\caption{Comparison of singing transcription from polyphonic music. The best score in each row is highlighted in bold.}
\label{table:sota}
\vspace{-2mm}
\end{table}

We finally compare our method to previous works in STP. The left side of Table \ref{table:sota} shows the accuracy metrics from recent works. HZ is a rule-based model submitted to MIREX2020 by Zhuang He and Yin Feng \cite{he2021singing}. VOCANO is a semi-supervised STP model based on virtual adversarial training \cite{hsu2021vocano}. EFN is a model based on EfficientNet and trained with MIR-ST500 \cite{wang2021preparation}. Note that VOCANO and EFN require singing voice separation as a pre-processing in the inference phase. The three previous works reported the results on Cmedia but the result on MIR-ST500 is available only for EFN because the dataset was released recently and EFN was introduced as a baseline.  Compared to our proposed method, the unsupervised model, JDC$_{note}$(U), achieves high accuracy than HZ and VOCANO in all metrics on the Cmedia test set. This validates that the proposed method is superior to the semi-supervised method in VOCANO or the rule-based approach in HZ. However, JDC$_{note}$(U) has lower accuracy than EFN and the gap is larger on the MIR-ST500 test set because EFN was trained with the same MIR-ST500 dataset (but the training split). For fairness, we can compare EFN to JDC$_{note}$(L). The result shows that JDC$_{note}$(L) is better than EFN in COnPOff and CONP on Cmedia but the result is reverse on MIR-ST500. Given that JDC$_{note}$(L) was also trained with the same MIR-ST500 training set, the two models seem to be comparable to each other. However, JDC$_{note}$(U+L) pushes the accuracy levels higher, achieving best performances in COnPOff and CoOnP on Cmedia and in COnP and CON on MIR-ST500.   

\section{Conclusions}
We presented a method for STP that uses pre-trained vocal pitch estimation models and unlabeled datasets. The method converts the frame-level pseudo labels to note-level and augments the label quality through self-training in the teacher-student framework. Through the ablation study, we showed that the model trained through the proposed method can achieve comparable results to the previous works, and with additional labeled data, it achieves better performance than the model trained with only labeled data. Since the test sets cover only Chinese music in this paper, we plan to evaluate the method on various genres of music in different cultural backgrounds as future work. 

\section{Acknowledgement}
This research is supported Year 2022 Copyright Technology R\&D Program by Ministry of Culture, Sports and Tourism and Korea Creative Content Agency(Project Name: Development of high-speed music search technology using deep learning, Project Number: CR202104004)

\vfill\pagebreak



\bibliographystyle{IEEE}

\end{document}